\documentclass[fleqn,twoside]{article}
\usepackage{espcrc2}

\usepackage{graphicx}

\newcommand{\bel}{\begin{equation}}
\newcommand{\ee}{\end{equation}}
\def\rys#1#2{\begin{figure}[h]
      \vskip 3mm
      \centerline{
      \includegraphics*[width=0.5 \textwidth]{#1}
      }
      \caption{#2}
      \vskip 3mm
      \end{figure}
      }

\newcommand{\AmS}{{\protect\the\textfont2
  A\kern-.1667em\lower.5ex\hbox{M}\kern-.125emS}}

\title{Modelling nuclear effects
in neutrino interactions in 1 GeV region}

\author{Jan T. Sobczyk\address[IFT]{Institute of Theoretical Physics, Wroclaw
University.\\
pl. M. Borna 9, 50-204 Wroclaw, Poland}
        \thanks{Supported by KBN grant 344/SPB/ICARUS/P-03/DZ211/2003-2005}}

\begin{document}

\begin{abstract}
We evaluate nuclear effects in neutrino reactions in a framework based
on a model proposed by Marteau with quasi-elastic and $\Delta$ production processes
treated together. Nuclear effects include RPA corrections and $\Delta$ width
modification in nuclear matter.
\vspace{1pc}
\end{abstract}

\maketitle

\section{INTRODUCTION}
Description of $\Delta$ excitation region in neutrino-nucleus
interactions is the most problematic ingredient in Monte Carlo
codes \cite{nuint}. Experiments provide cross sections with a
precision $\sim 25\%$ \cite{barish,radecky}. MC implementations
are based on a combination of Rein-Sehgal pion production model
\cite{reinsehgal} with Fermi gas model with different level of
sophistication in kinematical assumptions. The basic dynamical
rule is a factorization of interaction in two steps: (i)
neutrino-nucleon interaction and
(ii) re-interactions of outgoing particles inside nucleus.\\
In this contribution we wish to present computations done in a
framework of more ambitious theoretical scheme. Our model is based
on Marteau model \cite{marteau} but it includes several
modifications and also simplifications made in order to be able to
compare better its predictions with experimental data. Marteau
model describes on equal footing $\Delta$ excitation and
quasi-elastic processes. It is based on the non-relativistic Fermi
gas with RPA corrections due to contact interaction terms and
exchange of pions and $\rho$ mesons \cite{oddz}. Elementary
$2p-2h$ excitations are also included. A modification of $\Delta$
width in a nuclear matter is done using Oset's results
\cite{oset}. The model provides inclusive cross section for
quasi-elastic and $\Delta$ excitation reactions and also
contributions from several exclusive channels. The original model
is rather complicated as local density effects are taken into
account from the very beginning \cite{marteau_phd}.
\medskip
\\
Our modifications are:
\smallskip
\\
i) We avoid complications due to local density profile of nucleus
keeping a constant value of Fermi momentum $k_F=225\ MeV$. The
model becomes easier to handle and local density effects can be
included at the very end in the MC approach.
\smallskip
\\
ii) We adopt relativistic nucleons kinematics (we use relativistic
generalization of the Lindhard function).
\smallskip
\\
iii) We adopt original Oset results for the
$\Delta$ width in nuclear matter to specific kinematics of neutrino-nucleus
reaction.
\smallskip
\\
iv) We do not include $2p-2h$ part as it requires further study
\cite{marteau_private}.

\section{FERMI GAS}
The basic cross section formula is:

\bel{d^2\sigma\over dqd\nu}={G_F^2\cos\theta_c^2q\over 32\pi
E^2}L_{\mu\nu}H^{\mu\nu}. \ee
where
\bel L_{\mu\nu}=8(k'_{\mu}k_{\nu}+k_{\mu}k'_{\nu}-g_{\mu\nu}k\cdot k'
+i\epsilon_{\mu\nu\alpha\beta}k'^{\alpha}k^{\beta})\ee
is the leptonic tensor, $E$ is the neutrino energy, $k,k'$ denote
lepton initial and final four-momenta, $q^{\mu}=k^{\mu}-k'^{\mu}=(\nu , \vec q)$ is
four-momentum transfer. The hadronic tensor is of the form
\bel H^{\mu\nu}=H_{NN}^{\mu\nu}+H_{N\Delta}^{\mu\nu}+H_{\Delta N}^{\mu\nu}
+H_{\Delta\Delta}^{\mu\nu}\ee
We use the hadronic weak current
\rys{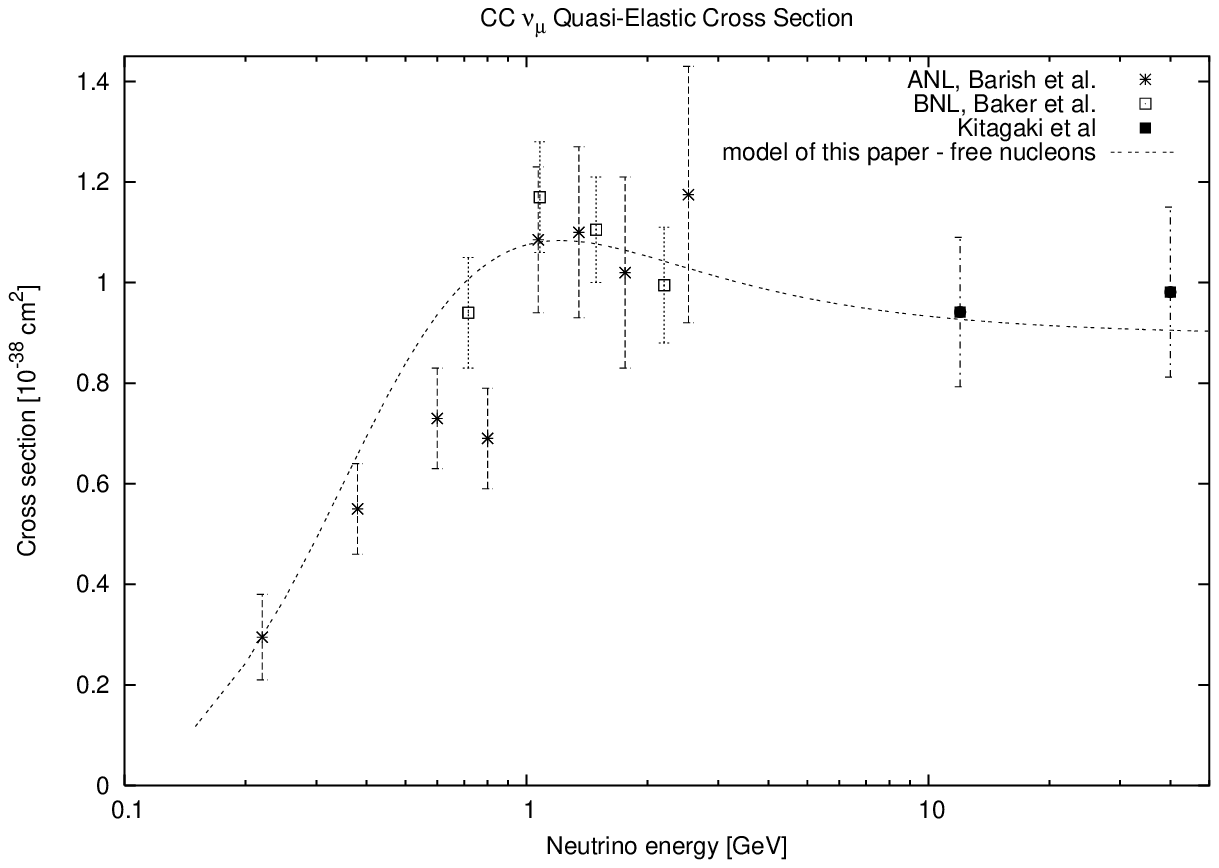}{CC Quasi-elastic $\nu_{\mu}$ cross section on free nucleons.
Experimental points are taken from \cite{barish2,baker,kitagaki}.}
\rys{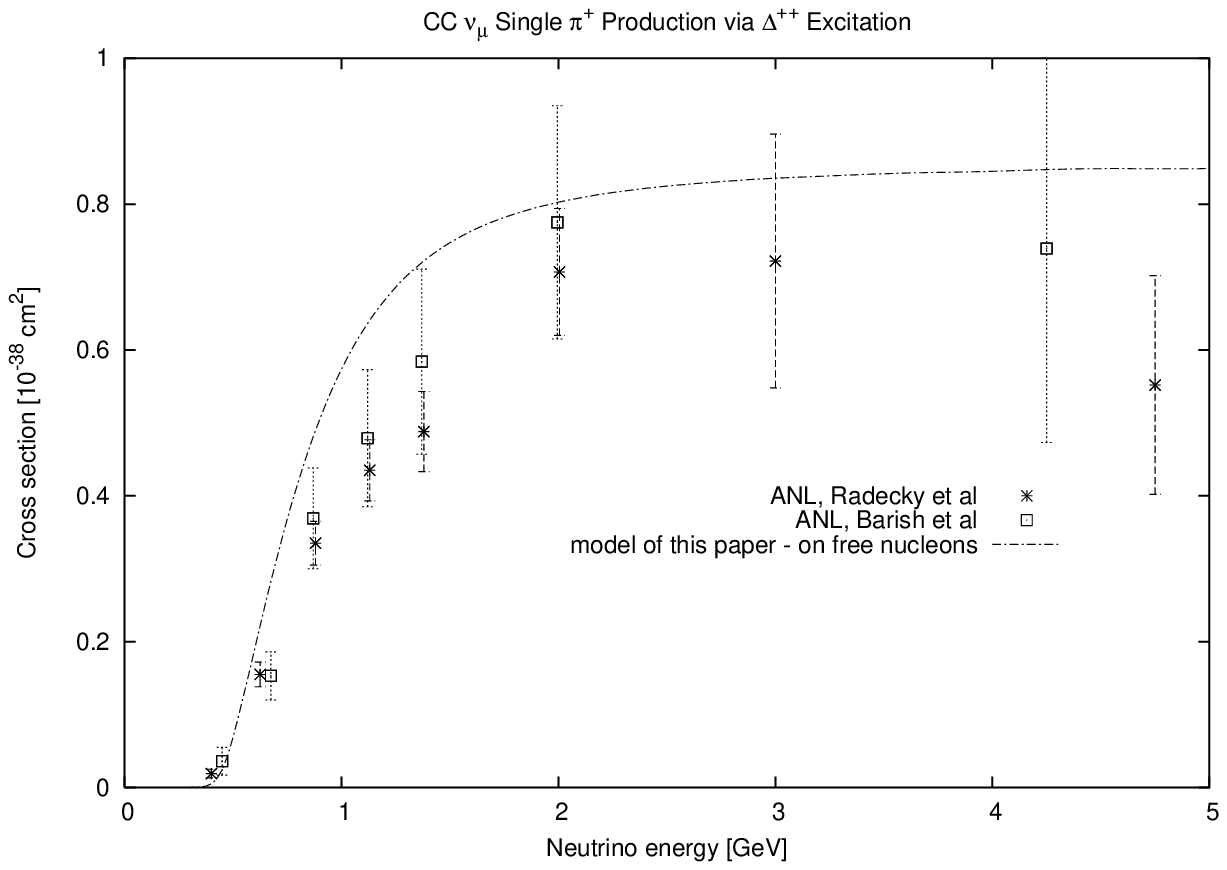}{CC $\Delta^{++}$ excitation $\nu_{\mu}$
cross section on free nucleons.
Experimental points are taken from \cite{barish,radecky}
and refer to $\nu_{\mu}p\rightarrow \mu\pi^+p$.}
\bel\begin{array}{l}\displaystyle
J^{\mu}=F_1(Q^2)\gamma^{\mu}+iF_2(Q^2)\sigma^{\mu\nu}{q_{\nu}\over
2M}\\[4mm]
\displaystyle \quad
+G_A(Q^2)\gamma^{\mu}\gamma_5+G_P(Q^2)\gamma_5{q^{\mu}\over
2M}\end{array}\ee where $F_1$, $F_2$, $G_A$ and $G_P$ are the
standard form-factors. In the frame $\vec q=(0,0,q)$ we calculate
($x,y\in \{N, \Delta\}$)\cite{marteau_phd}:
\bel\begin{array}{l}\displaystyle
H^{00}_{xy}(\nu ,q)=\sqrt{M_x+M+\nu\over 2M_x}\sqrt{M_y+M+\nu\over 2M_y}\\[4mm]
\displaystyle \qquad\qquad \times\Bigl(\alpha^0_{0x}(\nu
,q)\alpha^0_{0y}(\nu ,q)R^c_{xy}(\nu ,q)\\[4mm]
\displaystyle \qquad\qquad+\beta^0_{0x}(\nu ,q)\beta^0_{0y}(\nu
,q)R^l_{xy}(\nu ,q) \Bigr)\end{array}\ee
\bel \alpha^0_{0x}(\nu ,q)=F_1(Q^2)-F_2 (Q^2) {q^2\over
2M(M_x+M+\nu)}\ee
\bel\begin{array}{l}\displaystyle
\beta^0_{\ 0 x}(\nu ,q)=\\[4mm]
\displaystyle =q\Biggl( {G_A(Q^2)\over M_x+M+\nu} - {\nu\over
2M}\cdot {G_P(Q^2)\over M_x+M+\nu}\Biggr) ,\end{array}\ee etc.
The free Fermi gas is characterized by
\bel R_{N\Delta}=R_{\Delta N}=0,\ee
\bel R^{c,l,t}_{NN}(\nu ,q)=-{Vol\over \pi}{\cal
I}m\Pi^0_{N-h}(\nu ,q),\ee
\bel R^{l,t}_{\Delta\Delta}(\nu ,q)=-({f_{\pi N\Delta}\over f_{\pi
NN}})^2 {Vol\over \pi}{\cal I}m\Pi^0_{\Delta -h}(\nu ,q),\ee
\bel R^c_{\Delta\Delta}=0.\ee
\bel Vol={3\pi^2A\over 2k_F^3},\ee
\bel\begin{array}{l}\displaystyle {\cal I}m\Pi^0_{N-h}(\nu , \vec q)=
-{2M^2\over (2\pi )^2}\int d^3p\theta (k_F-|\vec p|)\\[4mm]
\displaystyle
\qquad\quad \times {\delta (\nu +
E_{\vec p}-E_{\vec q+\vec p})\over
E_{\vec p}E_{\vec q+\vec p}}
\theta (|\vec q+\vec p|-k_F)\end{array}\ee
\bel\begin{array}{l}\displaystyle {\cal I}m\Pi^0_{\Delta -h}(\nu , \vec q)=
\\[4mm]
\displaystyle
-{16\over 9}{M_{\Delta}^2\over (2\pi )^3}\int d^3p
{\Gamma_{\Delta}\cdot\theta (k_F-|\vec p|)\over
(s-M_{\Delta}^2)^2+
M_{\Delta}^2 \Gamma_{\Delta}^2}\end{array}\ee
\bel\Gamma_{\Delta}=PBL\cdot \Gamma_{\pi N}-2{\cal I}m(\Sigma_{\Delta})\ee
where $PBL\in [0,1]$ is the Pauli blocking factor defined as
follows \cite{oset}:
\rys{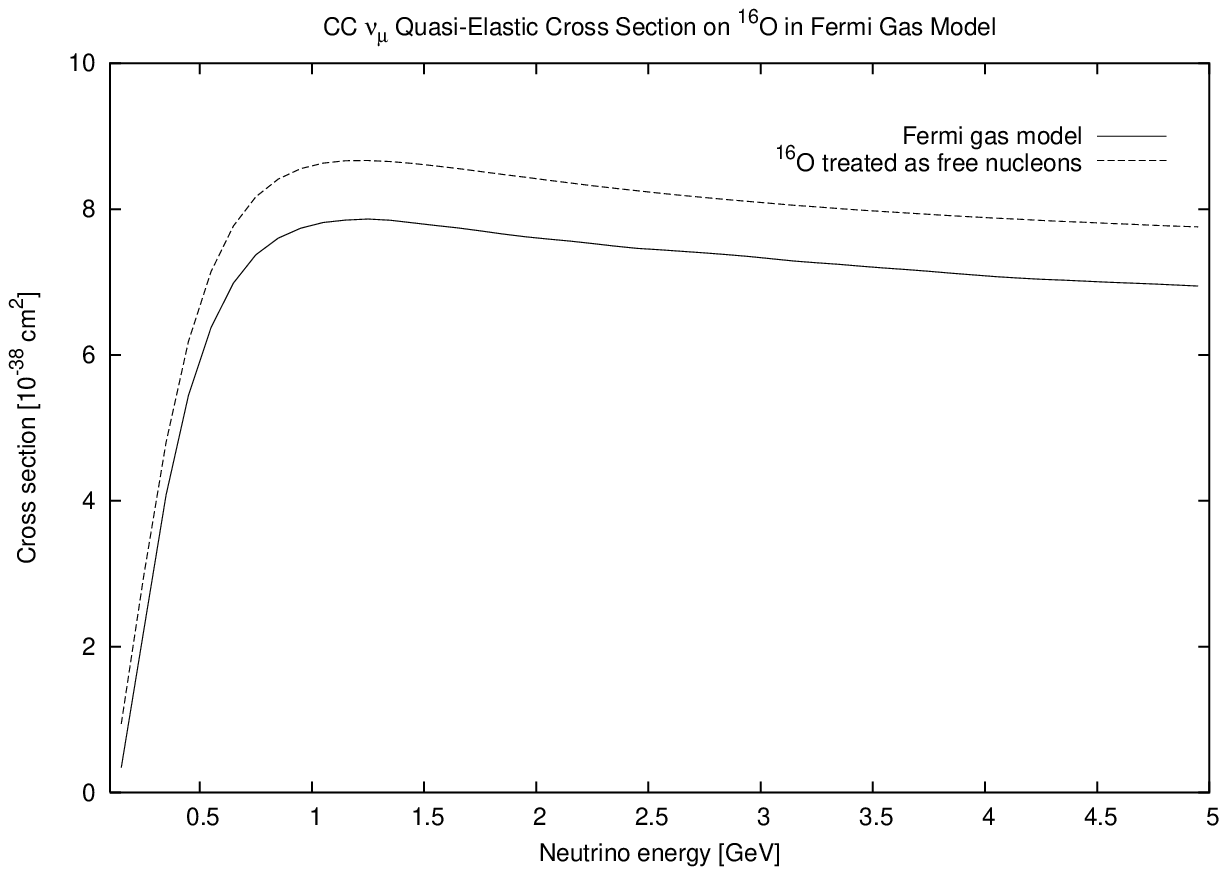}{CC Quasi-elastic $\nu_{\mu}$ cross section on $^{16}O$
in Fermi gas model.}
\rys{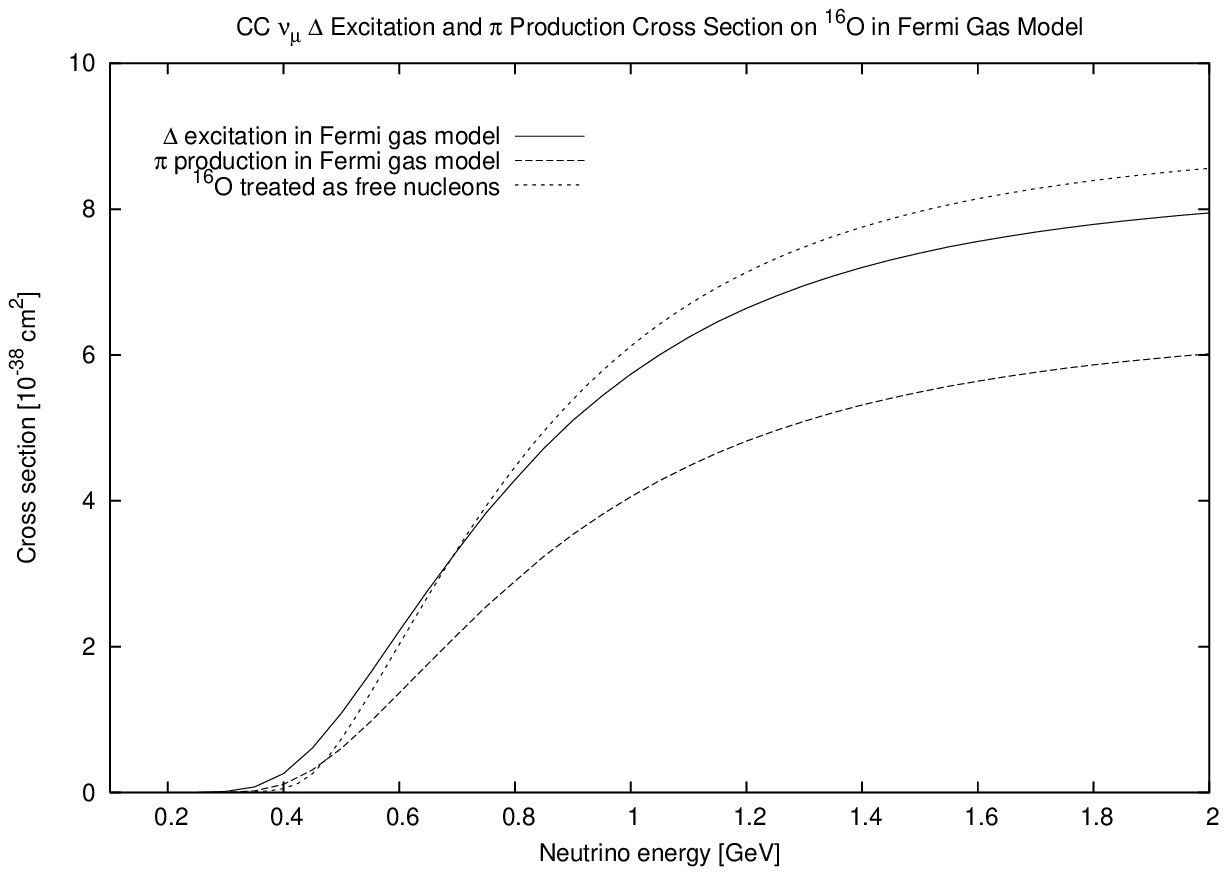}{CC $\nu_{\mu}$ $\pi$ production cross section on $^{16}O$
in Fermi gas model is smaller then $\Delta$ excitation cross section since
$\Delta$ can decay without $\pi$'s in final state}
\bel PB={|\vec p_{\Delta}||\vec q_{cm}|-\sqrt{s}E_F+E_{\Delta}E_{cm}
\over 2|\vec p_{\Delta}||\vec q_{cm}|} \ee
$PBL=1$ if $PB>1$, $PBL=0$ if $PB<0$, otherwise $PBL=PB$.
$(E_{\Delta}, \vec p_{\Delta})$ is
$\Delta$ $4-$momentum,
$(E_{cm}, \vec q_{cm})$ is pion (from $\Delta$ decay)
$4-$momentum in the center of mass frame,
\bel\Gamma_{\pi N}=\Gamma_0{q_{cm}(\sqrt{s})^3\over
q_{cm}(M_{\Delta})^3} {M_{\Delta}\over \sqrt{s}},\ee
$\Gamma_0=115MeV$, $M_{\Delta}=1232MeV$. ${\cal
I}m(\Sigma_{\Delta})$ describes nuclear effects in the form of
extra contributions to the $\Delta$ width from channels
$\Delta\rightarrow \pi N$, $N\Delta\rightarrow NN$,
$NN\Delta\rightarrow NNN$.
\bel {\cal I}m(\Sigma_{\Delta})={\cal I}m(\Sigma^{\pi}_{\Delta})+
{\cal I}m(\Sigma^{NN,NNN}_{\Delta}).\ee
${\cal I}m(\Sigma_{\Delta})$ was calculated by Oset-Salcedo \cite{oset} in two
kinematical situations: pion-nucleon and photon-nucleon
scatterings. Because the kinematical region for neutrino induced
reaction is different ($\nu^2-\vec q^2<0$) we adopt an
approximation
$${\cal I}m(\Sigma_{\Delta})_{\nu}={\cal I}m(\Sigma_{\Delta})_{\gamma}+
({\cal I}m(\Sigma_{\Delta})_{\gamma}-{\cal I}m(\Sigma_{\Delta})_{\pi})$$
\bel\ee
with ${\cal I}m(\Sigma_{\Delta})_{\gamma}$, ${\cal I}m(\Sigma_{\Delta})_{\pi}$
taken from Oset-Salcedo. Our prescription introduces an increase with respect
to the value of ${\cal I}m(\Sigma_{\Delta})_{\gamma}$ by amount of 5-10\%.

Normalization factors are checked by performing the limit
$k_F\rightarrow 0$ in which we recover quasi-elastic (Fig. 1)
and $\Delta$ excitation (Fig. 2)
cross sections on free nucleons. In the case of $\Delta$
excitation Marteau model provides a sum of cross sections over
isospin states i.e. a sum of  and $\Delta^+$ productions. We
recover $\Delta^{++}$ production cross section assuming isospin
branching ratio rule $\sigma (\Delta^{++})=3\sigma (\Delta^+)$. We
perform a comparison for this particular channel of pion
production since it is known that in the reaction $\nu_{\mu}
p\rightarrow \mu^- p \pi^+$ resonance contribution is dominant.
Experimental points are taken from papers \cite{barish2,baker,kitagaki}
in the case of quasi-elastic reaction and \cite{barish,radecky} in the case of $\Delta^{++}$ production.

In calculating $H^{\mu\nu}$ we adopted an approximation in which
calculation of sums over hadronic spins is done in the limit in
which target nucleon is at rest. In the case of quasi-elastic one
can verify by explicit computations that this approximation is
very good (on the level of 2-3\%). In the case of $\Delta$
excitation computations of response functions done by Marteau \cite{marteau_phd} show
that in the relevant kinematical region the approximation is
also very good.

Results in the Fermi gas model are presented in Fig. 3 (quasi-elastic reaction)
and Fig. 4 ($\Delta$ excitation and $\pi$ production). In these and next
computations we have assumed that
nucleus in question is $^{16}O$ i.e. $A=16$.
As mentioned in the introduction we do not take into account $^{16}O$
density profile keeping a constant value of Fermi momentum.

\section{RPA CORRECTIONS}

RPA equations read (separately for c,l,t; we omit the arguments
$\nu$ and $q$ of all the functions below):

\rys{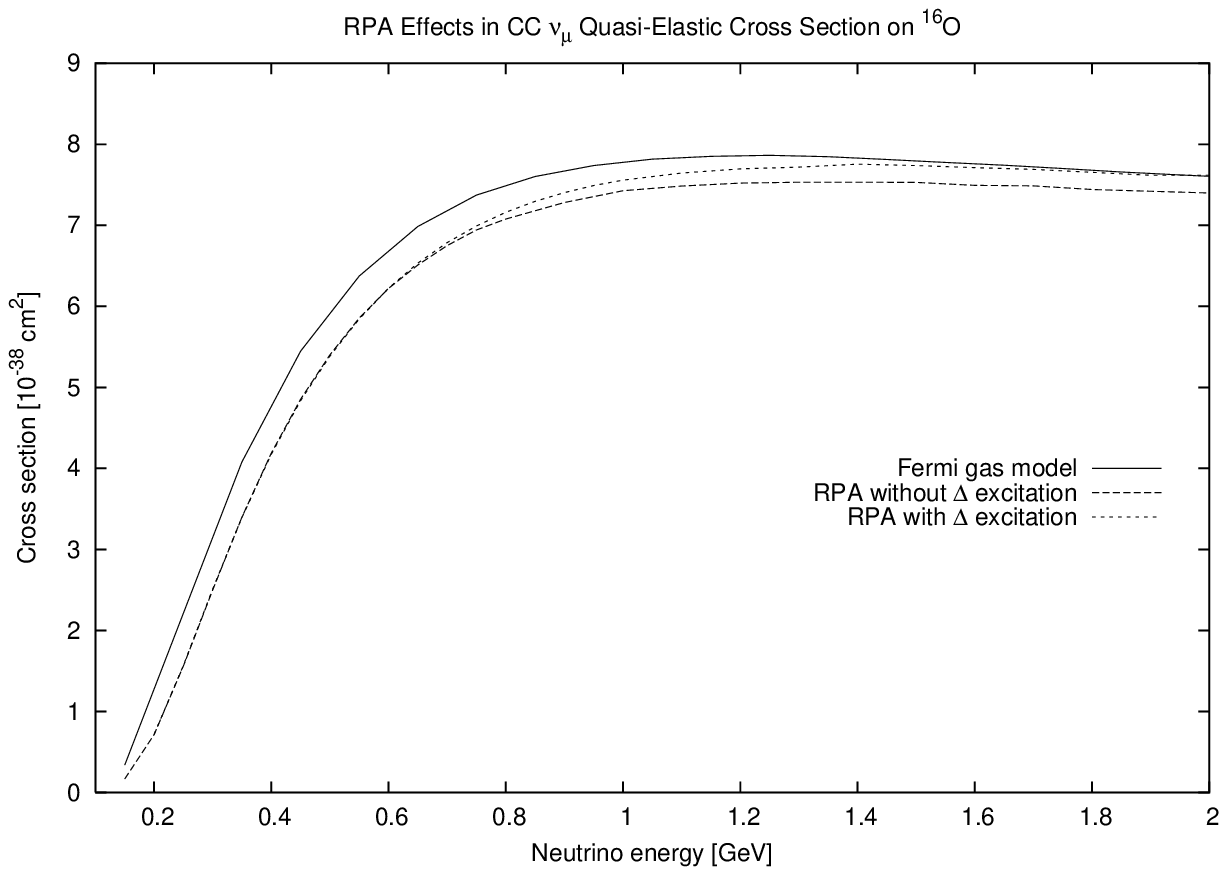}{CC $\nu_{\mu}$ quasi-elastic cross section on $^{16}O$ with
RPA corrections.}
\rys{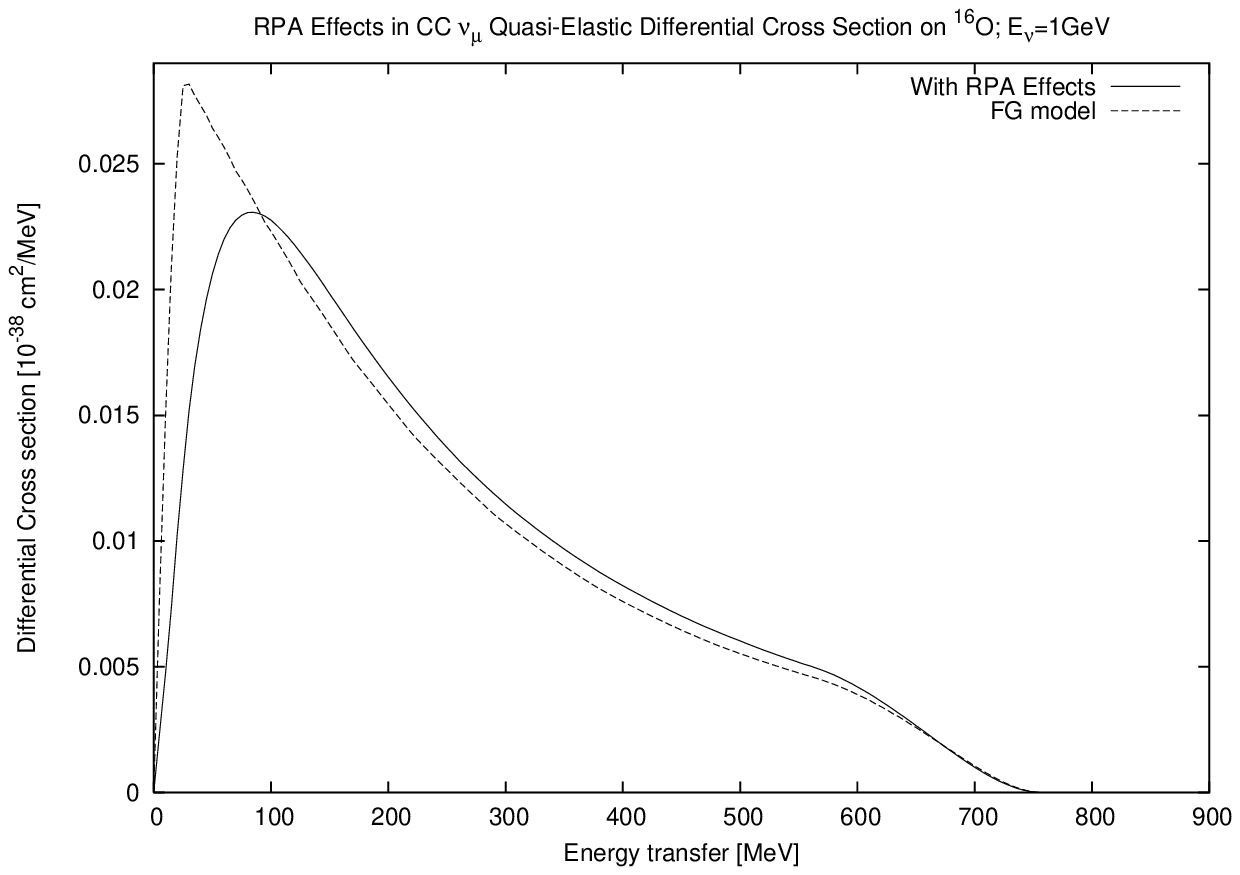}{CC $\nu_{\mu}$ differential
quasi-elastic cross section on $^{16}O$ with
RPA corrections; neutrino energy is $1GeV$.}
\rys{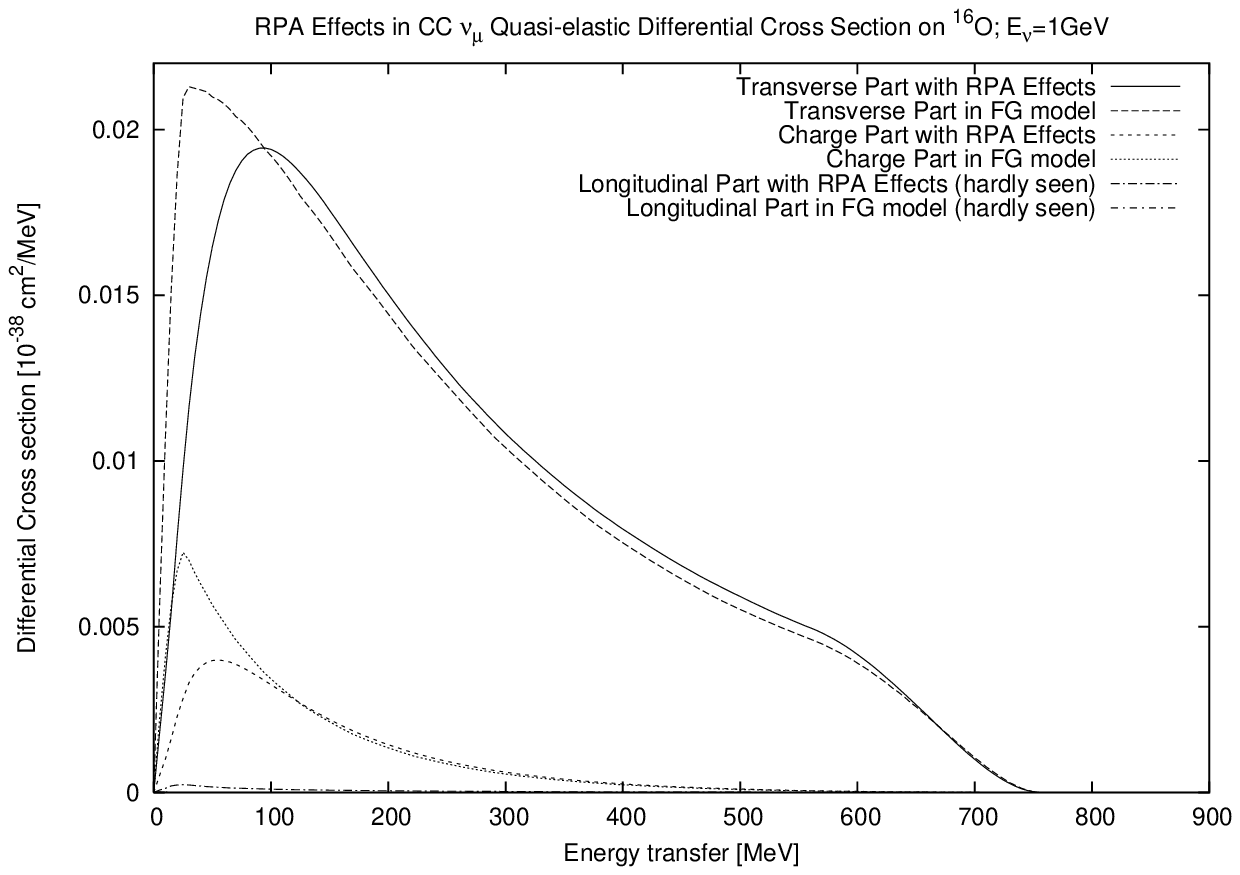}{CC $\nu_{\mu}$ differential
quasi-elastic cross section on $^{16}O$ with
RPA corrections in decomposition to charge, longitudinal and transverse
parts; neutrino energy is $1GeV$.}

$$\Pi_{NN}=\Pi^0_{N-h}+\Pi^0_{N-h}V_{NN}\Pi_{NN}+\Pi^0_{N-h}V_{N\Delta}
\Pi_{\Delta N},$$
$$\Pi_{\Delta\Delta}=\Pi^0_{\Delta -h}+\Pi^0_{\Delta -h}
V_{\Delta N}\Pi_{N\Delta}+\Pi^0_{\Delta -h}V_{\Delta\Delta}
\Pi_{\Delta \Delta},$$
\bel\begin{array}{l}\displaystyle
\Pi_{N\Delta}=\Pi^0_{N-h}V_{NN}\Pi_{N\Delta}+\Pi^0_{N-h}V_{N\Delta}
\Pi_{\Delta \Delta},\\[4mm]
\displaystyle
\Pi_{\Delta N}=\Pi^0_{\Delta -h}
V_{\Delta N}\Pi_{NN}+\Pi^0_{\Delta -h}V_{\Delta\Delta}
\Pi_{\Delta N}.\end{array}\ee
The solutions are found to be

\bel\begin{array}{l}\displaystyle
\Pi_{NN}=\Pi^0_{N-h}(1-V_{\Delta\Delta}\Pi^0_{\Delta -h}) D^{-1}\\[4mm]
\displaystyle
\Pi_{\Delta\Delta}=\Pi^0_{\Delta -h}(1-V_{NN}\Pi^0_{N-h}) D^{-1}\\[4mm]
\displaystyle
\Pi_{N\Delta}=\Pi_{\Delta N}=
V_{N\Delta}\Pi^0_{\Delta -h}\Pi^0_{N-h} D^{-1}
\end{array}
\ee
where
\bel\begin{array}{l}\displaystyle
D=(1-V_{NN}\Pi^0_{N-h})(1-V_{\Delta\Delta}\Pi^0_{N-h})\\[4mm]
\displaystyle
\qquad\qquad -V_{N\Delta}^2 \Pi^0_{N-h}\Pi^0_{\Delta -h}\end{array}\ee
After substitution

\bel R^{l,t}_{\Delta\Delta}=-({f_{\pi N\Delta}\over f_{\pi NN}})^2
{Vol\over\pi}{\cal I}m\Pi^{l,t}_{\Delta\Delta},\ee

\bel R^{c,l,t}_{NN}=-{Vol\over\pi}{\cal I}m\Pi^{c,l,t}_{NN},\ee

\bel R^{l,t}_{N\Delta}=R^{l,t}_{\Delta N}=
-{f_{\pi N\Delta}\over f_{\pi NN}}{Vol\over\pi}{\cal I}m\Pi^{l,t}_{N\Delta}\ee
we obtain the final expression for inclusive the cross section.
\medskip
\\
In numerical computations we use the interaction terms
\cite{oddz}:

\bel\begin{array}{l}\displaystyle
V_c^{NN}={f'\over m_{\pi}^2},
\\[4mm]
\displaystyle
V_l^{NN}={f_{\pi NN}^2\over m_{\pi}^2}
\Bigl({\Lambda_{\pi}^2-m_{\pi}^2\over \Lambda_{\pi}^2
-\nu^2+q^2}\Bigr)^2
\\
\displaystyle
\qquad\qquad \times\Bigl(g'+{q^2\over
\nu^2-q^2-m_{\pi}^2}\Bigr),
\\[4mm]
\displaystyle
V_t^{NN}={f_{\pi NN}^2\over
m_{\pi}^2}\Biggl(g'\Bigl({\Lambda_{\pi}^2-m_{\pi}^2\over
\Lambda_{\pi}^2-\nu^2+q^2}\Bigr)^2 \\
\displaystyle
+\qquad C_{\rho}^2{q^2\over \nu^2-q^2-m_{\varrho}^2}
\Bigl({\Lambda_{\varrho}^2-m_{\varrho}^2\over \Lambda_{\varrho}^2
-\nu^2+q^2}\Bigr)^2\Biggr),
\end{array}\ee

\bel\begin{array}{l}
\displaystyle
V_l^{N\Delta}=V_l^{\Delta N}={f_{\pi NN}f_{\pi N\Delta}\over
m_{\pi}^2}\\
\displaystyle
\times\Bigl({\Lambda_{\pi}^2-m_{\pi}^2\over \Lambda_{\pi}^2
-\nu^2+q^2}\Bigr)^2\Bigl( g''+{q^2\over
\nu^2-q^2-m_{\pi}^2}\Bigr)\\[4mm]
\displaystyle
V_t^{N\Delta}=V_t^{\Delta N}={f_{\pi NN}f_{\pi N\Delta}\over m_{\pi}^2}\\
\displaystyle
\qquad \times\Biggl(g''\Bigl({\Lambda_{\pi}^2-m_{\pi}^2\over
\Lambda_{\pi}^2-\nu^2+q^2}\Bigr)^2\\
\displaystyle
+C_{\rho}^2{q^2\over
\nu^2-q^2-m_{\varrho}^2}
\Bigl({\Lambda_{\varrho}^2-m_{\varrho}^2\over \Lambda_{\varrho}^2
-\nu^2+q^2}\Bigr)^2\Biggr),
\end{array}\ee

\bel\begin{array}{l}
\displaystyle
V_l^{\Delta\Delta}=
{f_{\pi N\Delta}^2\over m_{\pi}^2}
\Bigl({\Lambda_{\pi}^2-m_{\pi}^2\over \Lambda_{\pi}^2
-\nu^2+q^2}\Bigr)^2
\\
\displaystyle\qquad\quad
\times\Bigl(g'''+{q^2\over
\nu^2-q^2-m_{\pi}^2}\Bigr),
\\[4mm]
\displaystyle
V_t^{\Delta\Delta}={f_{\pi N\Delta}^2\over
m_{\pi}^2}\Biggl(g''' \Bigl({\Lambda_{\pi}^2-m_{\pi}^2\over
\Lambda_{\pi}^2-\nu^2+q^2}\Bigr)^2 \\
\displaystyle \quad +C_{\rho}^2{q^2\over
\nu^2-q^2-m_{\varrho}^2}\cdot
\Bigl({\Lambda_{\varrho}^2-m_{\varrho}^2\over \Lambda_{\varrho}^2
-\nu^2+q^2}\Bigr)^2\Biggr),
\end{array}
\end{equation}

\rys{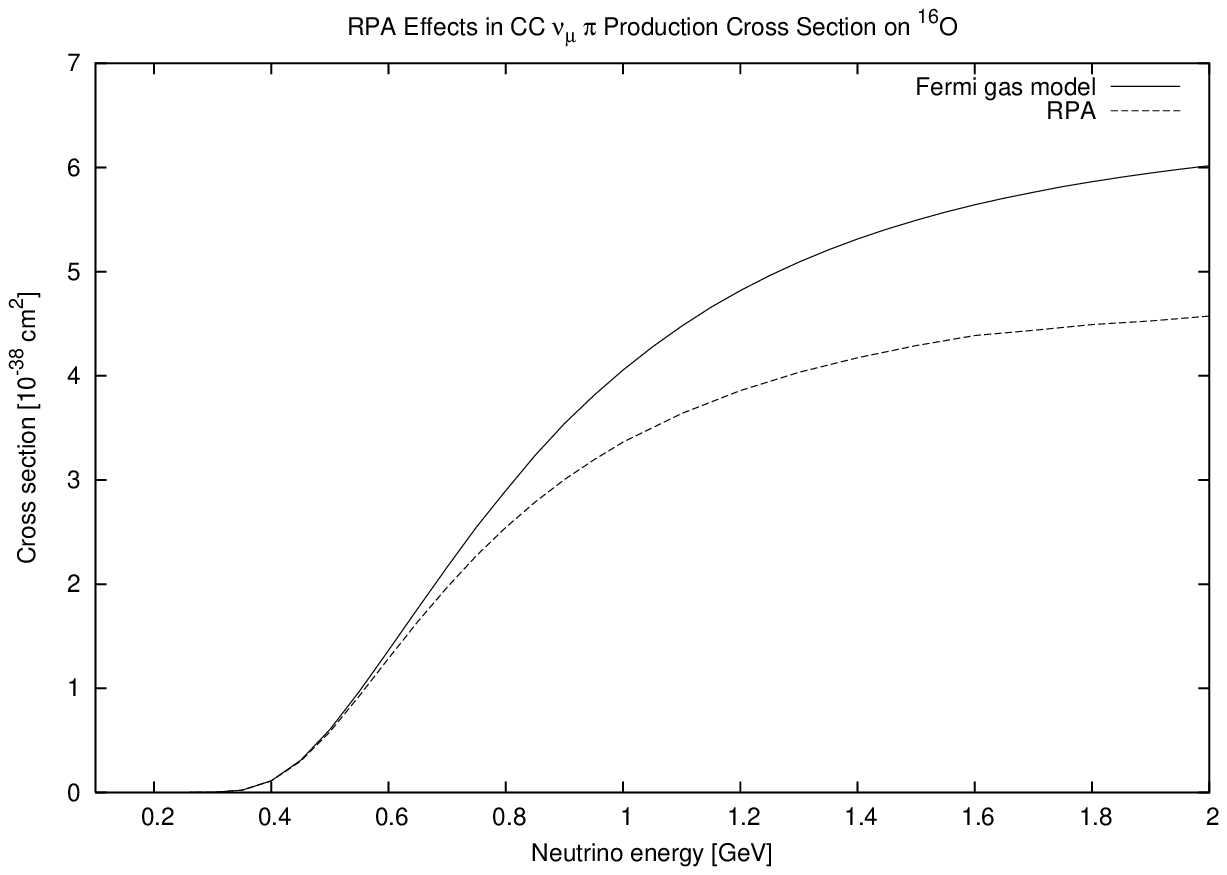}{$\nu_{\mu}$ CC $\pi$ production
cross section on $^{16}O$ with
RPA corrections.}
with the following values of free parameters:
\smallskip
\\
$f_{\pi NN}^2=4\pi\cdot 0.08$,
$({\displaystyle f_{\pi N\Delta}\over \displaystyle f_{\pi NN}})^2=4.78$,
$C_{\rho}^2=(
{\displaystyle m_{\pi}\over\displaystyle m_{\rho}}\cdot
{\displaystyle f_{\rho NN}\over\displaystyle f_{\pi NN}}
)^2=2$,
$f'=0.6$,
$g'=0.7$,
$g''=0.5$,
$g'''=0.5$,
$\Lambda_{\pi}=1000 MeV$,
$\Lambda_{\rho}=1500 MeV$.

\section{EXCLUSIVE CHANNELS}

In identifying contributions to exclusive channels we distinguish
quasi-elasic processes and reactions with and without pions in the final state.
For this
purpose we split $\Delta$ width and response functions in two parts according
to

\rys{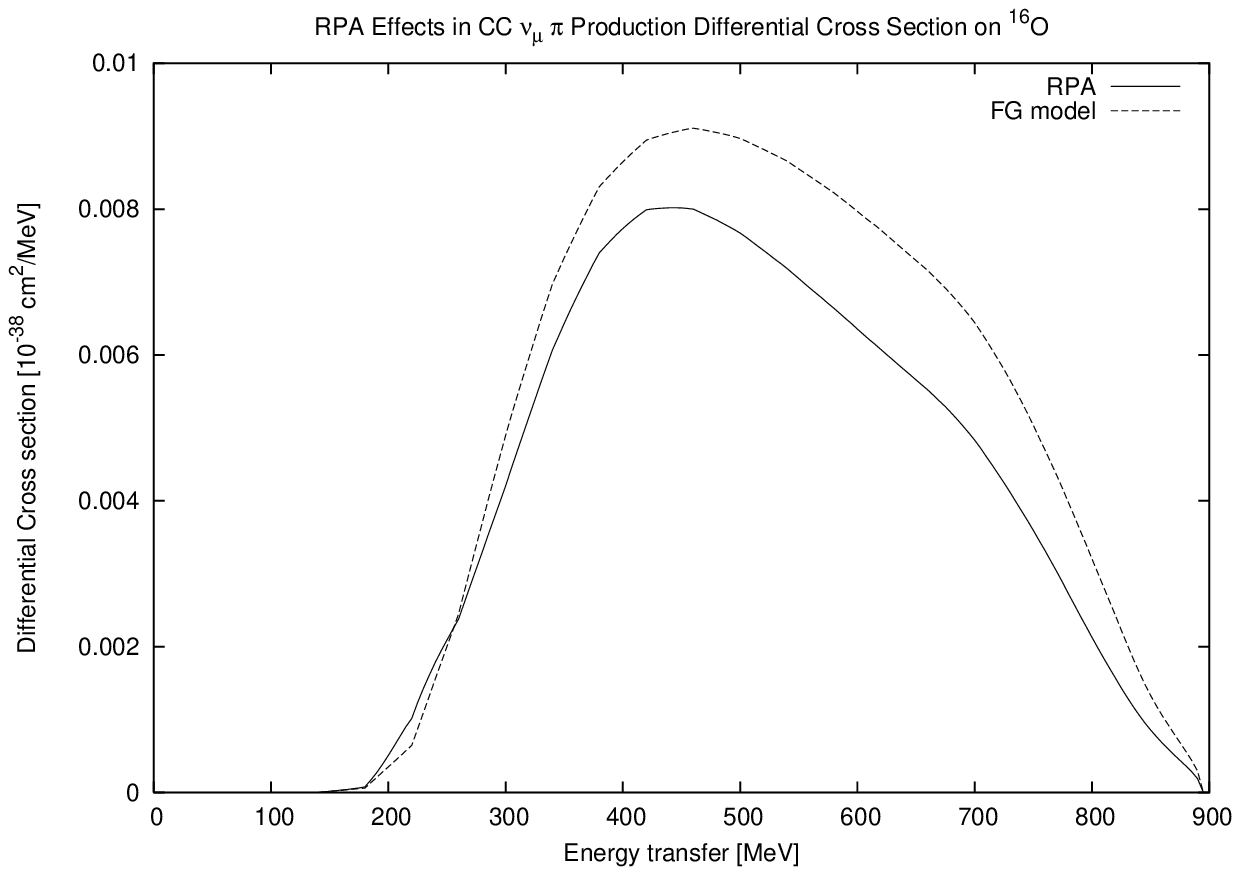}{$\nu_{\mu}$ CC differential
$\pi$ production cross section on $^{16}O$ with
RPA corrections; neutrino energy is $1GeV$.}
\rys{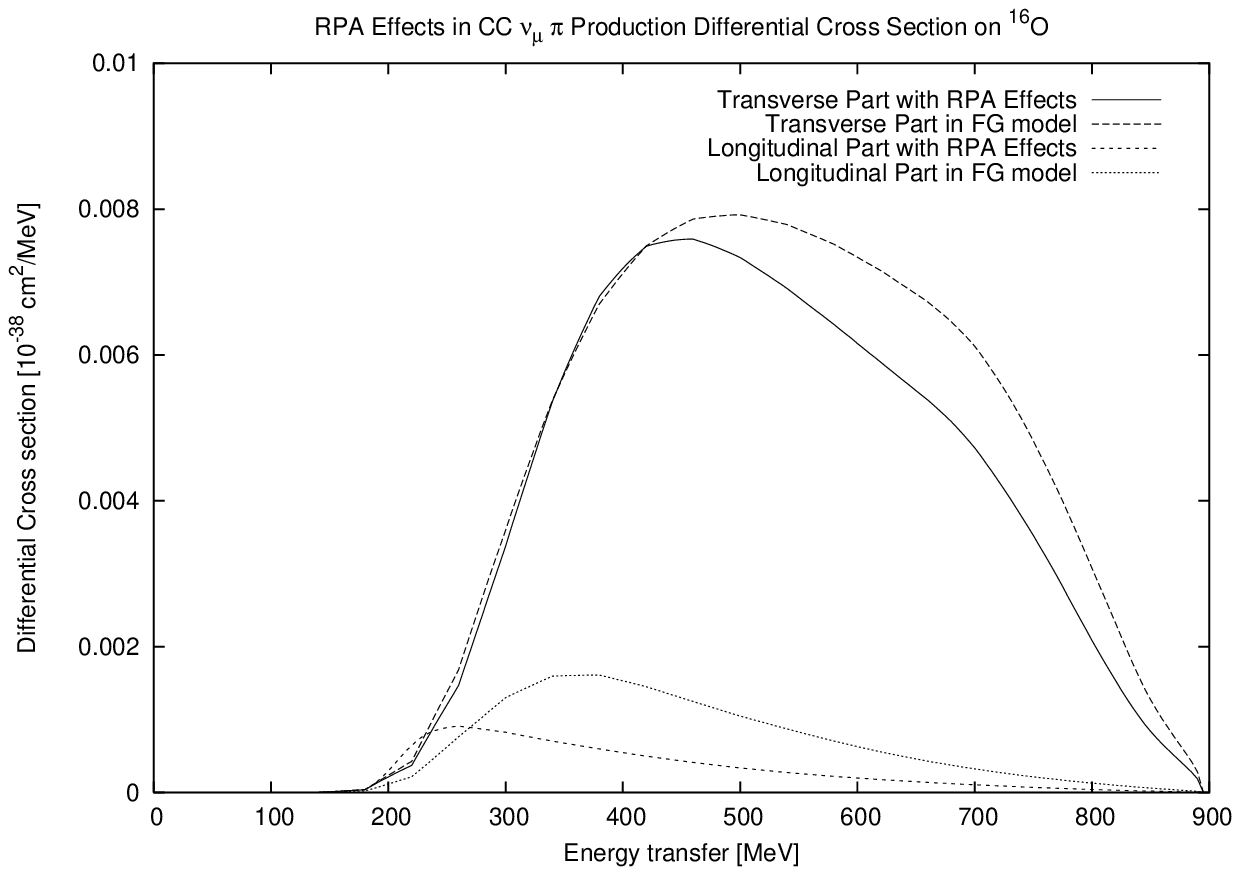}{$\nu_{\mu}$ CC differential
$\pi$ production cross section on $^{16}O$ with
RPA corrections in decomposition to longitudinal and transverse
parts; neutrino energy is $1GeV$.}
\bel\Gamma^{(\pi )}_{\Delta}=PBL\cdot \Gamma_{\pi N}-
2Im(\Sigma^{\pi}_{\Delta})\ee

\bel{\cal I}m\Pi^0_{\Delta -h}={\cal
I}m(\Pi^0_{\Delta -h})_{\pi}+ {\cal
I}m(\Pi^0_{\Delta -h})_{N}\ee

\bel \begin{array}{l}\displaystyle
{\cal I}m(\Pi^0_{\Delta -h})_{\pi}
(\nu , \vec q)=
\\[4mm]
\displaystyle -{16\over 9}{M_{\Delta}^2\over (2\pi )^3}\int d^3p
{\Gamma^{(\pi )}_{\Delta}\cdot\theta (k_F-|\vec p|)\over
(s-M_{\Delta}^2)^2+ M_{\Delta}^2 \Gamma_{\Delta}^2}\end{array}\ee
Contributions to pion production come from all four ${\cal
I}m\Pi_{xy}$ sectors of the theory. They are calculated by making
substitutions

\bel\begin{array}{l} \displaystyle
R^{l,t}_{N\Delta}\rightarrow
(R^{l,t}_{N\Delta})_{\pi}= -{f_{\pi N\Delta}\over f_{\pi
NN}}{Vol\over\pi}({\cal I}m\Pi^{l,t}_{N\Delta})_{\pi}\\[4mm]
\displaystyle
R^{l,t}_{NN}\rightarrow
(R^{l,t}_{NN})_{\pi}=-{Vol\over\pi}({\cal
I}m\Pi^{c,l,t}_{NN})_{\pi},\\[4mm]
\displaystyle
R^{l,t}_{\Delta\Delta}\rightarrow (R^{l,t}_{\Delta\Delta})_{\pi}
=-({f_{\pi N\Delta}\over f_{\pi NN}})^2{Vol\over\pi}({\cal
I}m\Pi^{l,t}_{\Delta\Delta})_{\pi},\end{array}\ee

\bel{\cal I}m(\Pi_{NN})_{\pi}= {|\Pi^0_{N-h}|^2(V_{N\Delta})^2
{\cal I}m(\Pi^0_{\Delta -h})_{\pi} \over |D|^2},\ee \bel \qquad
{1\over 2}({\cal I}m(\Pi_{N\Delta})_{\pi}+{\cal I}m(\Pi_{\Delta
N})_{\pi})=\ee
$${(V_{N\Delta}(Re(\Pi^0_{N-h})-V_{NN})|\Pi^0_{N-h}|^2)
{\cal I}m(\Pi^0_{\Delta -h})_{\pi} \over |D|^2},$$ \bel {\cal
I}m(\Pi_{\Delta\Delta})_{\pi}={|(1-V_{NN}\Pi^0_{N-h})|^2 {\cal
I}m(\Pi^0_{\Delta -h})_{\pi} \over |D|^2}\ee

\section{RESULTS AND DISCUSSION}

Our results concerning relevance of RPA are summarized in a series of plots.

In Fig.~5 total cross sections for CC quasi-elastic reaction are presented.
We compare three situations: Fermi gas without RPA correlations, RPA
correlations without taking into account $\Delta -h$ excitation and full RPA
computations. It is seen that RPA does not introduce much change in the
total cross section. Inclusion of $\Delta -h$ excitations increases
cross section
for neutrino energies above $0.8GeV$ and one arrives at results very close
to those for Fermi gas.

In Fig. 6 CC quasi-elastic differential cross section
${\displaystyle d\sigma\over \displaystyle
d\nu}$
is shown for neutrino energy $E=1GeV$. \\[1mm]
A well know result is reproduced:
the quasi-elastic peak is cut and some increase of differential
cross section is observed at
bigger values of energy transfer \cite{oddz}.

In Fig.~7  transverse, longitudinal and charge contributions
(according to spin operators present in the transition amplitude)
are distinguished in the CC quasi-elastic differential cross
section and patterns of RPA modifications for them is seen. The
transverse contribution dominates and the longitudinal one is
negligible.

In Fig. 8 total cross sections for CC $\pi$ production are
shown. We compare two situations: with and without RPA correlations.
Inclusion
of RPA correction reduces the cross section significantly by about 25\%.

In Fig. 9 CC $\pi$ production differential cross section
${\displaystyle d\sigma\over \displaystyle
d\nu}$ is shown for neutrino energy $E=1GeV$.
\vskip 1mm
In Fig. 10 in the CC $\pi$ production differential cross section
longitudinal and transverse contributions (according to spin
operators present in the transition amplitude) are distinguished
and patterns of RPA modifications can be followed. There is a
substantial reduction of the dominanting transverse part at bigger
values of energy transfer. The longitudinal part is in general
reduced (order of 50 \%) but some increase is also
seen for smaller values of
energy transfer.
\\
[2mm] To summarize: we have presented a scheme to calculate
nuclear effects in neutrino-nucleus interactions. As a candidate
to implement in MC codes the model presented in this paper has to
be supplemented with a non-resonant contribution to $\pi$
production, perhaps after \cite{foglinardulli}. It is unclear how
much of nuclear effects described by RPA and $\Delta$ width
modification are covered in FSI (Final State Interactions) models
of existing MC codes. This has to be understood in order
to avoid double counting. Finally, as mentioned in the introduction,
local density effects can be investigated by repeating present
computations for several values of Fermi momentum and by taking an
appropriate average.\\
[5mm] Acknowledgments: I thank D. Kielczewska for her encouraging
interest in this investigation and J. Marteau for useful
conversations at the early stage of work.

\end{document}